\documentclass{anstrans}
\title{Energetic Light Fragment Production Capability in MCNP6}
\author{Leslie M. Kerby,$^{*,\dagger}$ Stepan G. Mashnik,$^{\dagger}$ Konstantin K. Gudima,$^{\ddagger}$ Arnold J. Sierk,$^{\dagger}$ Jeffrey S. Bull,$^{\dagger}$ and Michael R. James$^{\dagger}$}

\institute{
$^{*}$Idaho State University, Pocatello, ID
\and
$^{\dagger}$Los Alamos National Laboratory, Los Alamos, NM
\and
$^{\ddagger}$Institute of Applied Physics, Academy of Science of Moldova, Chi\c{s}in\u{a}u, Moldova
}

\email{kerblesl@isu.edu}

\usepackage{graphicx} 
\usepackage{booktabs} 
\usepackage{microtype} 
\usepackage{enumitem}
\usepackage{gensymb}

\begin{document}
\section{Introduction}
The Monte Carlo Methods, Codes, and Applications group within the Computational Physics Division at Los Alamos National Laboratory has led the development of the transport code MCNP6 (Monte Carlo N-Particle transport code, version 6) \cite{MCNP6}. MCNP6 is a general-purpose, continuous-energy, generalized-geometry, time-dependent, Monte-Carlo radiation-transport code designed to track many particle types over broad ranges of energies. It is used around the world in applications ranging from radiation protection and dosimetry, nuclear-reactor design, nuclear criticality safety, detector design and analysis, decontamination and decommissioning, accelerator applications, medical physics, space research, and beyond. At lower energies, the code uses tables of evaluated nuclear data, while for higher energies ($>150$ MeV), MCNP6 uses the cascade-exciton model, version 03.03 (CEM03.03) \cite{CEMModel,ICTP-IAEAWorkshop}, and the Los Alamos quark-gluon string model, version 03.03 (LAQGSM03.03) \cite{ICTP-IAEAWorkshop, LAQGSM} to model nuclear reactions.

Emission of energetic heavy clusters heavier than $^4$He from nuclear reactions play a critical role in several applications, including electronics performance in space, human radiation dosages in space or other extreme radiation environments, proton- and hadron-therapy in medical physics, accelerator and shielding applications, and more. None of the available models are able to accurately predict emission of light fragments (LF) from arbitrary reactions. The CEM and LAQGSM event generators in MCNP6 describe quite well the spectra of fragments with sizes up to $^{4}$He across a broad range of target masses and incident energies (up to $\sim 5$~GeV for CEM and up to $\sim 1$~TeV/A for LAQGSM). However, they do not predict the high-energy tails of LF spectra heavier than $^4$He well. Most LF with energies above several tens of MeV are emitted during the precompound stage of a reaction. The 03.03 versions of CEM and LAQGSM do not account for precompound emission of LF larger than $^{4}$He. 

The goal of this research is to enable MCNP6 to produce high-energy light fragments. These energetic LF may be emitted by our models through three processes: Fermi breakup, preequilibrium, and coalescence. We explore the emission of light fragments through each of these mechanisms and demonstrate an improved agreement with experimental data achieved by extending precompound models to include emission of fragments heavier than $^4$He.

\section{Fermi Breakup}
The Fermi breakup model is used in CEM and LAQGSM for residual nuclei with atomic mass number $A \leq 12$, making it particularly important for reactions with light target nuclei.  

De-excitation of light nuclei with $A \leq A_{\rm Fermi}$ remaining after the Intra-Nuclear Cascade is described in CEM and LAQGSM only with the Fermi breakup model, where  $A_{\rm Fermi}$ is a `cut-off value' fixed in the models. The value of $A_{\rm Fermi}$ is a model parameter, not a physical characteristic of nuclear reactions. Actually, the initial version of the Fermi breakup model incorporated into CEM and LAQGSM used $A \leq A_{\rm Fermi} = 16$, just as $A_{\rm Fermi} = 16$ is used currently in GEANT4~\cite{GEANT4} and in SHIELD-HIT~\cite{SHIELD}. But that initial version of the Fermi breakup model had some problems and code crashes in some cases. To avoid unphysical results and code crashes, we chose the expedient of using $A_{\rm Fermi} = 12$ in both CEM and LAQGSM. Later, the problems in the Fermi breakup model were fixed, and we studied how the value of $A_{\rm Fermi}$ affected the final results of those codes.

We found that the performance of MCNP6, CEM, and LAQGSM in
simulating fragmentation reactions at intermediate energies for targets with
$A<13$ provide reasonably good
predictions for all reactions tested, although a fine-tuning of the $A_{Fermi}$ cut-off
parameter in the Fermi breakup model might provide a better description of some
experimental data. However, in some cases $A_{Fermi}=12$ provided the best agreement with experimental data, and so we did not find sufficient evidence to change the Fermi cut-off value in MCNP6, CEM, and LAQGSM at that time. See Ref.~\cite{NIMA2014} for complete results.

\section{Preequilibrium}
The preequilibrium interaction stage of nuclear reactions is considered by the
current CEM and LAQGSM in the framework of the latest version of the modified
exciton model (MEM)~\cite{Gudima,MODEX}.
At the preequilibrium stage of a reaction, CEM03.03 and LAQGSM03.03 take into
account all possible nuclear transitions changing the number of excitons $n$
with $\Delta_n = +2$, -2, and 0, as well as all possible multiple subsequent
emissions of n, p, d, t, $^3$He, and $^4$He. The corresponding system of master
equations describing the behavior of a nucleus at the preequilibrium stage is
solved by the Monte-Carlo technique. Improvements to the preequilibrium stage were three-fold: extension of preequilibrium to include LF emission up to $^{28}$Mg, adoption of the NASA reaction cross section as the inverse cross section, and the creation of a new model for the condensation probability $\gamma_j$.

\subsection{Preequilibrium extension}
CEM03.03 does not have the capability to output cross sections for fragments
larger than $^4$He. Therefore, one of the first things done was to add this
capability. We also created the capability to output by isotope, Z number, or
mass number. 

Extending the MEM to produce 66 fragment types, up to $^{28}$Mg,
involves extending the calculation of emission widths (Eq.~(\ref{GammaLambda})) to all
66 fragment types. The emission width $\Gamma _{j}$,
(or probability of emitting particle fragment $j$), is estimated as
\begin{equation}
\Gamma_{j}(p,h,E) = \int_{V_j^c}^{E-B_j} \lambda_j (p,h,E,T)dT ,
\label{GammaLambda}
\end{equation}
where the partial transmission probabilities, $\lambda_j$, are equal to
\begin{equation}
\begin{split}
\lambda_{j}(p,h,E,T) = & \gamma_j \frac{2s_j + 1}{\pi^2\hbar^3} \mu_j \Re (p,h)
 \frac{\omega (p_j,0,T+B_j)}{g_j} \\
& \times \frac{\omega (p-p_j,h,E-B_j-T)}{\omega (p,h,E)} T \sigma_j^{inv} (T) \mbox{ ,}
\end{split}
\label{eq:lambda_j}
\end{equation}
for complex particles and fragments. This extension therefore entails calculating Coulomb barriers, binding energies,
reduced masses, inverse cross sections, and condensation probabilities for all
66 fragment types. 

The extended MEM provides dramatically improved ability to describe
light-fragment production at intermediate to high energies across most reactions
tested, while maintaining good results for fragments $\leq ^4$He. 

\subsection{Inverse Cross Sections}
Total-reaction-cross-section models have a significant impact on the predictions
and accuracy of spallation and transport codes. For example, CEM uses total reaction cross sections as {\it inverse} cross sections, $\sigma^{inv}_j$,
to calculate the probabilities of emission of possible nucleons and
fragments.  

The inverse cross sections in CEM03.03 are taken from the Dostrovsky et al.
model \cite{Dostrovsky}. Its general form is:
\begin{equation}
\sigma_{Dost.} = \pi r_0^2 A^{2/3} \alpha_j(1 - \frac{V_j}{T}) .
\label{eq:Dost}
\end{equation}
The Dostrovsky model was not intended for use above about 50 MeV/nucleon,
and is not very suitable for emission of fragments heavier than $^4$He. Better
total-reaction-cross-section models are available today. We compared results from several total reaction cross section models and determined the NASA (or Tripathi et al.) model~\cite{NASAp} provided the best general agreement with experimental data. As shown in Eq.~(\ref{eq:NASA}), the NASA cross section attempts to simulate several
quantum-mechanical effects, such as the optical potential for neutrons
(with the parameter $X_m$) and collective effects like Pauli blocking (through
the quantity $\delta_T$).  
\begin{equation}
\sigma_{NASA} = \pi r^2_0 (A_P^{1/3} + A_T^{1/3} + \delta_T )^2 
(1 - R_c \frac{B_T}{T_{cm}})X_m \mbox{ .}
\label{eq:NASA}
\end{equation}

Results of implementing the NASA inverse-cross-section model into the extended MEM show improved
agreement with experimental data. See Ref.~\cite{NIMB2015} for details.

\subsection{Condensation probability $\gamma_j$}
The condensation probability, $\gamma_j$, represents the
probability that $p_j$ excited nucleons (excitons) will condense to form
a complex particle of type $j$ in the excited residual nucleus. We explore the formulation of a new model for $\gamma_j$, one which is
energy-dependent, and which provides improved fits to experimental
fragment spectra.

The condensation probability $\gamma_j$ could be calculated from first principles,
but such a calculation is not feasible. $\gamma_j$ is, therefore, estimated as the overlap integral of the
wave function of independent nucleons with that of the complex particle (see
details in~\cite{CEMModel}), as shown in Eq.~(\ref{GammaBeta}).

\begin{equation}
\gamma_j \approx p_j^3 (\frac{p_j}{A})^{p_j - 1} .
\label{GammaBeta}
\end{equation}

Eq.~(\ref{GammaBeta}) is a rather crude estimate. In CEM we approximate
$\gamma_j$ by multiplying the estimate provided by Eq.~(\ref{GammaBeta}) by
empirical coefficients $F_j$. Values of $F_j$ for d, t, $^3$He, and $^4$He need to be re-fit after the upgrades
to the inverse-cross-section and coalescence models, and new values of $F_j$ need
to be obtained for heavy clusters up to $^{28}$Mg. 

In analyzing the fitted $F_j$ data set we created mathematical models for both the fragment-specific equations for $F_j$ and a generalized $F_j$ model. The general $F_j$ model is:  
\begin{equation}
\begin{split}
\label{Fj_gen}
F_j(T_0,A_j,Z_j,A_t) = & \biggl[ 7800 (2.5)^{A_j} e^{-T_0/20} + 
 \frac{2 (4)^\tau}{T_0^{0.2\tau} + 100} \biggr] \\
	& \times e^{-\frac{300-A_t}{100}}, \\
	& \tau = A_j - (Z_j - 3). 
\end{split}
\end{equation}
For the fragment-specific $F_j$, discussion of the physical meaning of the model, and results, see Ref.~\cite{CPC2015}.

\section{Coalescence} 
CEM03.03 is
capable of producing light fragments up to $^4$He in its coalescence model.
We extend the coalescence model to be able to produce up to $^7$Be in 
CEM and up to $^{12}$C in LAQGSM.

When the cascade stage of a reaction is completed, CEM uses the coalescence model to create high-energy
d, t, $^3$He, and $^4$He fragments by final-state interactions among emitted
cascade nucleons outside of the target nucleus. The magnitude of the momentum, $p$, of each cascade nucleon is calculated
relativistically from its kinetic energy, $T$. We
assume that all the cascade nucleons having differences
in their momenta smaller than $p_c$ and with the correct
isotopic content form an appropriate composite particle. 

The coalescence model first checks all nucleons to form 2-nucleon pairs, their
momenta permitting. It then takes these 2-nucleon pairs and the single nucleons left and forms $^4$He, $^3$He, and/or tritium, their momenta permitting. The extended coalescence model further takes these two-nucleon pairs, tritium, $^3$He, and $^4$He to see if they can coalesce to form
heavier clusters: $^6$He, $^6$Li, $^7$Li or $^7$Be.

All coalesced nucleons are removed from the distributions of nucleons
so that atomic and mass numbers are conserved. 

Results show significant improvement in the production of heavy clusters across the energy range. However, too many alpha particles were lost (coalesced into heavy clusters);
so $p_c(^4{\mbox He})$ was increased to compensate. See Refs.~\cite{NUFRA2015slides} and \cite{ANS2015} for details. The new values for $p_c$ for the extended coalescence model are:
\begin{eqnarray}
p_c(d) & = & 90 \mbox{ MeV/c ;} \nonumber \\
p_c(t) & = & p_c(^3{\mbox He}) = 108 \mbox{ MeV/c ;} \\
p_c(^4{\mbox He}) & = & 130 \mbox{ MeV/c ;} \nonumber \\
p_c(LF) & = & 175 \mbox{ MeV/c .} \nonumber 
\label{eq:exp_coalescence1}
\end{eqnarray}
For $300$ MeV $< T < 1000$ MeV they are:
\begin{eqnarray}
p_c(d) & = & 150 \mbox{ MeV/c ;} \nonumber \\
p_c(t) & = & p_c(^3{\mbox He}) = 175 \mbox{ MeV/c ;} \\
p_c(^4{\mbox He}) & = & 205 \mbox{ MeV/c ;} \nonumber \\
p_c(LF) & = & 250 \mbox{ MeV/c .} \nonumber 
\label{eq:exp_coalescence2}
\end{eqnarray}

\section{MCNP6 Implementation}
The culmination of this work is the implementation of CEM with these heavy-cluster upgrades, which we call CEM03.03F, into MCNP6. 

The GENXS option allows for various cross sections to be tallied in MCNP6. Previously, production cross sections were only available for fragments up
to $^4$He. Thus, a necessary first step in implementing the improved CEM03.03F into MCNP6 involves extending the ability to output production cross
sections to heavy clusters. This GENXS upgrade accomplishes this and includes
the ability to tally and output double differential cross sections, 
angle-integrated or energy-integrated cross sections, as well as the total production cross section, for any heavy ion with valid ZAID. For details see Ref.~\cite{PHYSOR2016}.

CEM03.03F was implemented into a working version of MCNP6, which we call MCNP6-F. Two of the upgrades are
always implemented: the upgraded NASA-Kalbach inverse
cross sections in the preequilibruim stage, and the new energy-dependent
$\gamma_j$ model. The other two upgrades (extension of preequilibrium emission to $^{28}$Mg, and the extension of the coalescence model to $^7$Be), both of
which increase computation time, may be turned off if desired. A variable, called
{\sf npreqtyp}, was created to specify the number of preequilibrium particles
considered for emission. It is now the twelfth option on the LCA card 
of the MCNP6 input file. Its
maximum (and default) value is 66, similar to the {\sf nevtype} variable used
for the evaporation stage. See Ref.~\cite{RMP2016} for a list of the 66
particles considered in the preequilibrium stage. In the old model, 6
preequilibrium particles were considered, and therefore a value of 
{\sf npreqtyp}=6 turns off both the preequilibrium and coalescence extensions.
The extended coalescence model is implemented for values of {\sf npreqtyp}$>$6.
MCNP6-F also includes the GENXS extension.

Basic testing and verification of MCNP6-F has been completed with the results
being presented below. In addition, MPI testing has been
completed. Upon further testing, we anticipate these heavy-ion upgrades and the
GENXS extension will be included in the next release of MCNP6.

\subsection{Results}

\begin{figure}[t!]
\centering
\includegraphics[width=3.5in]{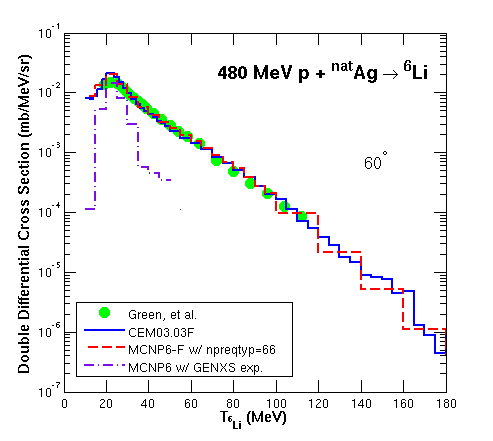}
\caption[MCNP6-F results for 480 MeV p + $^{nat}$Ag $\rightarrow$ $^6$Li]
 {Comparison of experimental data for 480 MeV p + $^{nat}$Ag $\rightarrow$
 $^6$Li at 60\degree, measured by Green, et al.\ \cite{Green480} (green
 circles) to calculated results from CEM03.03F (blue solid lines), MCNP6-F
 with {\sf npreqtyp}=66 (red dashed lines), and MCNP6 with the GENXS extension
 only (purple dash-dotted lines).}
\label{p480AgMCNP}
\end{figure}

Double differential cross section spectra for several reactions are plotted in this section. Fig.~\ref{p480AgMCNP} shows the results for 480 MeV p + $^{nat}$Ag $\rightarrow$ $^6$Li at 60\degree, compared to experimental data measured by Green, et al.~\cite{Green480}. 
MCNP6 with the GENXS extension only does not contain any of the four
light-fragment upgrades discussed in this work. MCNP6-F produces significantly improved results and matches the data reasonably well.

\begin{figure}[t!]
\centering
\includegraphics[width=3.5in]{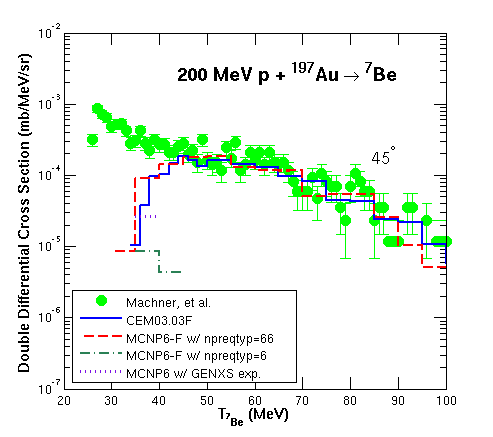}
\caption[MCNP6-F results for 200 MeV p + $^{197}$Au $\rightarrow$ $^7$Be]
 {Comparison of experimental data for 200 MeV p + $^{197}$Au $\rightarrow$
 $^7$Be at 45\degree, measured by Machner, et al.\ \cite{Machner} (green
 circles) to calculations from CEM03.03F (blue solid lines), MCNP6-F with 
 {\sf npreqtyp}=66 (red dashed lines), MCNP6-F with {\sf npreqtyp}=6 (green
 dash-dotted lines), and MCNP6 with the GENXS extension only (purple dotted
 lines).}
\label{p200AuMCNP}
\end{figure}

Fig.~\ref{p200AuMCNP} displays the results for 200 MeV p + $^{197}$Au $\rightarrow$ $^7$Be at 45\degree, compared to experimental data by Machner, et al.~\cite{Machner}. 
 This figure shows not only dramatically improved heavy-cluster production by MCNP6-F at high energies, but also improved production at relatively low energies around the
preequilibrium peak. We believe this is due to the heavy target (gold) and
therefore an increased ability to produce these low-energy heavy clusters from
both the extended coalescence model and the extended preequilibrium model.

Across the reactions we tested we found that the new MCNP6-F, in general, gives improved results
compared to the unmodified MCNP6, most especially for heavy-cluster spectra. Further details and more results can be found in Ref.~\cite{RMP2016}.
 
The goal of producing energetic light fragments in MCNP6 has been accomplished by implementing several heavy-cluster upgrades as outlined in this paper.

\section{Acknowledgments}
We are grateful to Drs. Christopher Werner and Avneet Sood of Los Alamos National Laboratory for encouraging discussions and support.

This study was carried out under the auspices of the National Nuclear Security Administration of the U.S. Department of Energy at Los Alamos National Laboratory under Contract No. DE-AC52-06NA25396.

This work was supported in part (for L.M.K) by the M. Hildred Blewett Fellowship of the American Physical Society, 
{\noindent
www.aps.org.
}

\bibliographystyle{ans}
\bibliography{bibliography}

\begin{thebibliography}{10}
\newcommand{\enquote}[1]{``#1''}

\bibitem{MCNP6}
\MakeUppercase{T.~Goorley}, et al., \enquote{Initial {MCNP6} Release Overview,
  {MCNP6} version 0.1,} \emph{Nuclear Technology}, \textbf{180}, 298--315
  (2012).

\bibitem{CEMModel}
\MakeUppercase{K.~Gudima}, \MakeUppercase{S.~Mashnik}, and
  \MakeUppercase{V.~Toneev}, \enquote{Cascade-Exciton Model of Nuclear
  Reactions,} \emph{Nuclear Physics A}, \textbf{401}, 329--361 (1983).

\bibitem{ICTP-IAEAWorkshop}
\MakeUppercase{S.~Mashnik}, \MakeUppercase{K.~Gudima},
  \MakeUppercase{R.~Prael}, \MakeUppercase{A.~Sierk},
  \MakeUppercase{M.~Baznat}, and \MakeUppercase{N.~Mokhov}, \enquote{{CEM03.03
  and LAQGSM03.03} Event Generators for the {MCNP6, MCNPX, and MARS15}
  Transport Codes,} Joint {ICTP-IAEA} Advanced Workshop on Model Codes for
  Spallation Reactions. Trieste, Italy (February 2008), {LANL} Report
  LA-UR-08-2931, arXiv:0812.1820.

\bibitem{LAQGSM}
\MakeUppercase{K.~Gudima}, \MakeUppercase{S.~Mashnik}, and
  \MakeUppercase{A.~Sierk}, \enquote{User Manual for the code {LAQGSM},}
  (2001), {LANL Report LA-UR-01-6804;
  http://lib-www.lanl.gov/lapubs/00818645.pdf}.

\bibitem{GEANT4}
\MakeUppercase{I.~Pshenichnov}, \MakeUppercase{A.~Botvina},
  \MakeUppercase{I.~Mishustin}, and \MakeUppercase{W.~Grainer},
  \enquote{Nuclear Fragmentation Reactions in Extended Media Studied with
  GEANT4 Toolkit,} \emph{Nuclear Instruments and Methods in Physics Research
  B}, \textbf{268}, 604 (2010).

\bibitem{SHIELD}
\MakeUppercase{M.~Hultqvist}, \MakeUppercase{M.~Lazzeroni},
  \MakeUppercase{A.~Botvina}, \MakeUppercase{I.~Gudowska},
  \MakeUppercase{N.~Sobolevsky}, and \MakeUppercase{A.~Brahme},
  \enquote{Evaluation of Nuclear Reaction Cross-Sections and Fragment Yields in
  Carbon Beams Using the SHIELD-HIT Monte Carlo Code. Comparison with
  Experiments,} \emph{Physics in Medicine \& Biology}, \textbf{57}, 4369
  (2012).

\bibitem{NIMA2014}
\MakeUppercase{S.~Mashnik} and \MakeUppercase{L.~Kerby}, \enquote{MCNP6
  Fragmentation of Light Nuclei at Intermediate Energies,} \emph{Nuclear
  Instruments and Methods in Physics Research A}, \textbf{764}, 59 (2014),
  arXiv:1404.7820.

\bibitem{Gudima}
\MakeUppercase{K.~Gudima}, \MakeUppercase{G.~Ososkov}, and
  \MakeUppercase{V.~Toneev}, \enquote{Model for Pre-Equilibrium Decay of
  Excited Nuclei,} \emph{Yadernaya Fizika}, \textbf{21} (1975), [Soviet Journal
  of Nuclear Physics 21 (1975) 139-143].

\bibitem{MODEX}
\MakeUppercase{S.~Mashnik} and \MakeUppercase{V.~Toneev}, \enquote{MODEX--the
  Program for Calculation of the Energy Spectra of Particles Emitted in the
  Reactions of Pre-Equilibrium and Equilibrium Statstical Decays,} \emph{JINR
  Communication}, \textbf{P4-8417} (1974).

\bibitem{Dostrovsky}
\MakeUppercase{I.~Dostrovsky}, \MakeUppercase{Z.~Fraenkel}, and
  \MakeUppercase{G.~Friedlander}, \enquote{Monte Carlo Calculations of Nuclear
  Evaporation Processes. III. Applications to Low-Energy Reactions,}
  \emph{Physical Review}, \textbf{116}, 683 (1959).

\bibitem{NASAp}
\MakeUppercase{R.~Tripathi}, \MakeUppercase{F.~Cucinotta}, and
  \MakeUppercase{J.~Wilson}, \enquote{Accurate Universal Parameterization of
  Absorption Cross Sections,} \emph{Nuclear Instruments and Methods in Physics
  Research B}, \textbf{117}, 347 (1996).

\bibitem{NIMB2015}
\MakeUppercase{L.~Kerby} and \MakeUppercase{S.~Mashnik}, \enquote{Total
  Reaction Cross Sections in CEM and MCNP6 at Intermediate Energies,}
  \emph{Nuclear Instruments and Methods B}, \textbf{356-357}, 135 (2015).

\bibitem{CPC2015}
\MakeUppercase{L.~Kerby}, \enquote{An Energy-Dependent Numerical Model for the
  Condensation Probability, $\gamma_j$,} LANL Report, LA-UR-15-26648 (2015).

\bibitem{NUFRA2015slides}
\MakeUppercase{K.~Gudima}, \MakeUppercase{S.~Mashnik}, and
  \MakeUppercase{L.~Kerby}, \enquote{Fragmentation of Light Nuclei at
  Intermediate Energies Simulated with {MCNP6},} LANL Report, LA-UR-15-27417,
  presented at the Fifth International Conference on Nuclear Fragmentation From
  Basic Research to Applications (NUFRA2015), 4 -- 11 October 2015, Kemer
  (Antalya), Turkey (2015).

\bibitem{ANS2015}
\MakeUppercase{L.~Kerby} and \MakeUppercase{S.~Mashnik}, \enquote{Production of
  Heavy Clusters with an Expanded Coalescence Model in CEM,} \emph{Transactions
  of the American Nuclear Society}, \textbf{112}, 577 (2015).

\bibitem{PHYSOR2016}
\MakeUppercase{L.~Kerby}, \MakeUppercase{S.~Mashnik}, and
  \MakeUppercase{J.~Bull}, \enquote{{MCNP6 GENXS} Option Expansion to Include
  Fragment Spectra of Heavy Ions,} LANL Report, LA-UR-15-27858 (2015), summary
  accepted for presentation at PHYSOR 2016, Sun Valley, Idaho, USA, May 1-5,
  2016.

\bibitem{RMP2016}
\MakeUppercase{L.~Kerby}, \MakeUppercase{S.~Mashnik},
  \MakeUppercase{K.~Gudima}, \MakeUppercase{A.~Sierk}, \MakeUppercase{J.~Bull},
  and \MakeUppercase{M.~James}, \enquote{Production of Energetic Heavy Clusters
  in {CEM and MCNP6},} LANL Report, LA-UR-15-29524 (2015).

\bibitem{Green480}
\MakeUppercase{R.~Green}, \MakeUppercase{R.~Korteling}, and
  \MakeUppercase{K.~Jackson}, \enquote{Inclusive Production of Isotopically
  Resolved Li Through Mg Fragments by 480 {MeV} p+Ag Reactions,} \emph{Physical
  Review C}, \textbf{29}, 1806--1824 (1984).

\bibitem{Machner}
\MakeUppercase{H.~Machner}, et al., \enquote{Isotopic production cross sections in
  proton-nucleus collisions at 200 {MeV},} \emph{Physical Review C},
  \textbf{73}, 044606 (2006).

\end{thebibliography}
\end{document}